\documentclass[prd,twocolumn,showpacs,tightenlines,%
superscriptaddress,nofootinbib,notitlepage,preprintnumbers]{revtex4-1}

\usepackage{graphicx}
\usepackage{latexsym}
\usepackage{amssymb}
\usepackage{amsmath}
\usepackage{color}
\usepackage{hyperref}

\begin{document}

\title{Black Hole in a Radiation-Dominated Universe}
\author{E.  Babichev}
\affiliation{Laboratoire de Physique Theorique, Universite Paris-Saclay, Paris, France}
\author{V. Dokuchaev}
\affiliation{Institute for Nuclear Research, Russian Academy of Sciences,
pr. 60-Letiya Oktyabrya 7a, Moscow, 117312 Russia}
\affiliation{MEPhI National Research Nuclear University, Kashirskoe sh. 31, Moscow, 115409 Russia}
\author{Yu. Eroshenko}
\affiliation{Institute for Nuclear Research, Russian Academy of Sciences,
pr. 60-Letiya Oktyabrya 7a, Moscow, 117312 Russia}

\date{\today}

\begin{abstract}
We study a black hole in an expanding Universe during the radiation-dominated stage. In the case when the black hole radius is much smaller than the cosmological horizon, we present a solution of the Einstein equations for the metric, the matter density and velocity distributions. 
At distances much smaller than the cosmological horizon the solution features quasi-stationary accretion, while at large distances the solution asymptotes the homogeneous radiation dominated FLRW. We discuss how our solution is related to the McVittie solution. 
The obtained results can be applied, in particular, for the formation of dark matter density spikes around primordial black
holes, and for the evolution of dark matter clumps during the radiation-dominated stage.
\end{abstract}

\maketitle 




\section{INTRODUCTION}

The metric of a black hole (BH) in an expending universe has been discussed since 1933,
starting with the paper by McVittie (1933) (for historical introduction and references
see Faraoni and Jacques 2007). McVitte postulated an solution in isotropic coordinates for a
BH in an expanding Universe, assuming absence of matter (accretion) flow. 
However, the McVittie metric has several drawback. In particular,  it has been shown that 
when matter consists of a perfect fluid with the energy-momentum tensor $T_{\mu\nu}=(p+\rho)u_\mu u_\nu-pg_{\mu\nu}$,
the generalised McVittie metric cannot be a solution of the Einstein equations (Faraoni and Jacques 2007),
except for the case of Schwarzschild-de-Sitter metric, realised for $p=-\rho$. 
The problems of the McVittie solution are due to the assumption of zero matter flow, which 
is clearly unphysical, since near a black hole a fluid should be attracted towards the black hole, therefore creating a non-zero flux.
Previously there were attempts to generalize the McVittie solution
to include radial accretion flow. This can be done, e.g.,  by introducing a mixture of two fluids:
usual dust and ``null dust'', a gas of massless particles (Sultana and Dyer 2005).
Another type of solutions can be obtained for imperfect fluid with a radial heat flow
onto a BH (Faraoni and Jacques 2007; McClure and Dyer 2006).
Instead of perfect fluid, one can pose a similar problem for a black hole in FLRW filled with a scalar field,
e.g., such a study of scalar field accretion onto a BH in an almost
de Sitter Universe was done in by Chadburn and
Gregory (2014).

In this paper we address the problem of a BH in
a radiation-dominated Universe from a different perspective.
We do not aim at an exact solution, instead
we look for an approximate  solution
that consistently describes the metric as well as the density and velocity distributions of a perfect fluid
at all ranges from the BH horizon 
to the cosmological one. We will consider the situation when
the cosmological radius is much greater than
the BH radius. In this case there are
small parameters in the problem, and one can look for
a solution as an expansion in terms of these parameters.
Following this approach, we find an approximate
solution of the Einstein equations: we present the BH metric in a radiation dominated
Universe in curvature coordinates, as well as density velocity distribution of the fluid. 
In contrast
to the McVittie solution, our solution can be applied
to physically relevant problems, such as physics of cosmological primordial BHs, or structure of dark matter density spikes
around primordial BHs (Eroshenko 2016). For the latter,
the distance where the fluid velocity changes sign (the turning point), calculated in this paper, 
is particularly important. 

The formation of primordial black holes was
considered in many papers using numerical methods
(Nadezhin et al. 1978; Novikov et al. 1979; Bicknell and Henriksen 1979; Polnarev and Musco 2007;
Musco et al. 2009; Polnarev et al. 2012). However,
in most cases, the numerical integration stops shortly
after the formation of the event horizon. In contrast,
in this paper we consider later times (after the BH
formation), when acoustic waves produced by the
collapse are scattered away and a steady quasi-state
phase is established. It should be noted that our approach
allows one to find a solution for times that are
orders of magnitude greater than the BH formation
time.
In this sense our work is complementary to the above mentioned studies of primordial BH formation.

To match the BH and the cosmological metrics,
it is useful to bring the two metrics in a similar
form. Such a method has been applied using
the Schwarzschild--de-Sitter coordinates, e.g., in
Babichev and Esposito--Farese (2013). In this paper
as common coordinates we choose the curvature
coordinates $(t,r,\theta,\phi)$ (for more details on the classification
of coordinate systems, see Bronnikov and Rubin
2008), in which the Schwarzschild metric is initially
written. The transformation of the cosmological
metric to curvature coordinates in a general form was
considered by Stanyukovich and Shershekeev (1966)
for three (open, flat, and closed) cosmological models
and for any time dependence of the scale factor
of the Universe. In this paper we show that for
a radiation-dominated Universe the corresponding
transformation can be easily derived in a different way
and written as a closed-form expression.
We then use this form of the cosmological metric to find a matching with the BH metric.


\section{THE COSMOLOGICAL METRIC IN CURVATURE COORDINATES}

Consider the following transformation of the conformal
cosmological metric to the metric in curvature
coordinates:
\begin{equation}
ds^2 = a^2(\eta)[d\eta^2-d\chi^2-\chi^2d\Omega^2]=e^{\nu}dt^2 - e^{\lambda}dr^2 -r^2d\Omega^2,
\label{EF2}
\end{equation}
where $\eta$ is the conformal time, and we use the units
$c=G=1$. Let us consider a BH at the stage of radiation domination.
Then,
\begin{equation}
\label{aeta}
a(\eta)=\eta,
\end{equation}
where the coefficient of proportionality is chosen to be unity for simplicity (this can always be done by
the redefinition of time).

We will look for the change of variables
\begin{equation}
t=t(\xi), \quad r=r(\eta,\chi),\quad   \mbox{ãäå} \quad \xi=\frac{\eta^2+\chi^2}{2},\nonumber
\end{equation}
that relates the left and right parts in Eq.~(\ref{EF2}). Denoting
the derivatives by subscripts, and substituting
\begin{equation}
dt=t_\xi \eta d\eta+t_\xi \chi d\chi, \quad dr=r_\eta d\eta+r_\chi d\chi,
\nonumber
\end{equation}
into (\ref{EF2}) and comparing similar terms, we obtain the
system of equations
\begin{equation}
\label{eq12}
\begin{split}
e^\nu \eta^2t_\xi^2-e^\lambda \chi^2 =\eta^2, \\
e^\nu \chi^2t_\xi^2-e^\lambda \eta^2=-\eta^2, \\
e^\nu t_\xi^2\chi \eta-e^\lambda r_\eta r_\chi=0,\\
r^2=\eta^2\chi^2
\end{split}
\end{equation}
where we used (\ref{aeta}). From the last equation we immediately
derive
\begin{equation}
r=a\chi=\eta\chi,
\label{retachi2}
\end{equation}
therefore, $r_\eta=\chi$ and $r_\chi=\eta$. 
In addition, it is easy to show only three equations out of four in the above system~(\ref{eq12}) are independent. 
The solution can be written as
\begin{equation}
e^\nu=\frac{\eta^2}{2t_\xi^2(\eta^2-\xi)}, \quad  e^\lambda=\frac{\eta^2}{2(\eta^2-\xi)}.  
\label{sol12}
\nonumber
\end{equation}

Further, note that the function $t(\xi)$ so far remains arbitrary
and, for simplicity we set
\begin{equation}
t(\xi)=\xi, \label{vxixi2}
\end{equation}
From (\ref{retachi2}) and (\ref{vxixi2}) we then derive
\begin{equation}
\eta^2=2\xi-\chi^2=2t-\frac{r^2}{\eta^2},
\label{quadr2}
\nonumber
\end{equation}
Hence we find a solution for $\eta$ in terms of $t$ and $r$:
\begin{equation}
\eta^2(t,r)=t + \sqrt{t^2-r^2},
\label{etafun}
\end{equation}
where we chose the positive sign in front of the
square root, since the solution with the minus sign
corresponds to a contracting Universe. As a result,
we have
\begin{equation}
\label{sol12finup}
e^\nu=e^\lambda=\frac{1}{2}\left(1+\frac{t}{\sqrt{t^2-r^2}}\right). 
\end{equation}
Note that $e^\nu$ and $e^\lambda$ coincide, giving the flat two-dimensional
part $dt^2-dr^2$. In contrast to the standard conformally flat form of FLRW metric,
the factor in front $dt^2-dr^2$ is radius-dependent.

Now let us express the cosmological density
\begin{equation}
\rho_{\infty}(\eta)=\frac{3}{8\pi \eta^4}
\nonumber
\end{equation}
in coordinates $(t,r)$. Using (\ref{etafun}), we obtain
\begin{equation}
\rho_{\infty}(t,r)=\frac{3}
{8\pi \left(t +\sqrt{t^2-r^2}\right)^2}.
\label{rhoinfty2}
\end{equation}
Note that the density in coordinates $(t,r)$ is not uniform
at $t=const$, because these hypersurfaces do not
coincide with the hypersurfaces $\eta=const$.

In the cosmological coordinates $(\eta,\chi)$ the 4-velocity components
of the fluid in the Universe are $u^0=1/a$ and $u^1=0$. Transforming this 4-velocity to the coordinate
system $(t,r)$, we get
\begin{equation}
u^0=1, \quad u\equiv u^1=\frac{r}{t+\sqrt{t^2-r^2}}.
\label{ucosm}
\end{equation}

Note that the obtained expressions  are only applicable for the region
inside the cosmological horizon $r<t$, similar to the case of de-Sitter solution, 
where only a patch of the space-time is covered by static coordinates.


\section{MATCHING THE BH METRIC AND THE COSMOLOGICAL METRIC}

Consider the range of radii
\begin{equation}
r_g\ll r\ll t,
\label{region1}
\end{equation}
i.e., distances much greater than the BH radius, but much smaller than the cosmological horizon size,
by assuming that a sufficiently long time has
passed since the BH formation and the cosmological horizon has
expanded to $t\gg r_g$ (We recall that primordial
BHs are formed at $t\sim r_g$).
First, note that in the
range (\ref{region1}) both BH and cosmological metric coefficients
can be written as $e^\nu=1+\varepsilon$, where $\varepsilon\ll1$, therefore, $\nu\approx\varepsilon$ is a small quantity,
and similarly for $\lambda$. Second, note that, in this case,
the Einstein equations can be linearized in $\nu$ and
$\lambda$. Indeed, the Einstein equations read (Landau and
Lifshitz 1975)
\begin{eqnarray}
 \label{G01}
 8\pi T^1_0 &=& -e^{-\lambda}\frac{\dot\lambda}{r}, \\
 \label{G00}
 8\pi T^0_0 &=& -e^{-\lambda}\left(\frac{1}{r^2}-
 \frac{\lambda'}{r}\right)+\frac{1}{r^2}, \nonumber\\
 \label{G11}
 8\pi T^1_1 &=& -e^{-\lambda}\left(\frac{1}{r^2}+
 \frac{\nu'}{r}\right)+\frac{1}{r^2}, \nonumber\\
 \label{G22}
 8\pi T^2_2 &=& \frac{e^{-\nu}}{2}
\left[\ddot\lambda+\frac{\dot\lambda}{2}\left(\dot\lambda-\dot\nu\right)\right] \nonumber\\
 &&-\frac{e^{-\lambda}}{2}
 \left[\nu''+(\nu'-\lambda')\left(\frac{\nu'}{2}+\frac{1}{r}\right)\right].\nonumber
\end{eqnarray}
While the energy-momentum tensor of a perfect fluid is
\begin{eqnarray}
 \label{T01}
  T^1_0 &=&
 (\rho+p)u\sqrt{\frac{f_0}{f_1}(f_1+u^2)}, \nonumber\\
 \label{T00}
 T^0_0 &=& \rho+(\rho+p)\frac{u^2}{f_1},  \nonumber\\
 \label{T11}
 T^1_1 &=& -\left[(\rho+p)\frac{u^2}{f_1}+p\right], \nonumber\\
 \label{T22}
 T^2_2 &=& -p,\nonumber
 \end{eqnarray}
where we denoted $f_0=e^{\nu}$ and $f_1=e^{-\lambda}$. 
For $\nu\ll1$, $\lambda\ll1$, and $u\ll1$ we can perform linearization and retain only the the first powers of $\nu$ and $\lambda$:
\begin{eqnarray}
 \label{G01L}
 8\pi (\rho+p)u &=& -\frac{\dot\lambda}{r}, \\
 \label{G00L}
 8\pi \rho &=& \frac{\lambda'}{r}+\frac{\lambda}{r^2}, \nonumber\\
 \label{G11L}
 -8\pi p &=& -\frac{\nu'}{r}+\frac{\lambda}{r^2}, \nonumber\\
 \label{G22L}
 -8\pi p &=& \frac{\ddot\lambda}{2}-\frac{\nu''}{2}-\frac{\nu'-\lambda'}{2r}.\nonumber
\end{eqnarray}
In addition, note that for a Schwarzschild BH $\rho=p=0$ everywhere, except for the singularity, and the
Schwarzschild solution for the metric
\begin{eqnarray}
\begin{aligned}
e^\nu  & =  1-\frac{r_g}{r},
\label{nuS}\\
e^\lambda  & =  \left(1-\frac{r_g}{r}\right)^{-1},
\end{aligned}
\end{eqnarray}
identically nullifies the right-hand side of the equations.
This means that in the range (\ref{region1}) the solution
for the metric coefficients in the first order is simply
the sum of the solutions for the BH and cosmology.
Expanding (\ref{sol12finup}) in terms of the small parameter $r/t$,
we obtain
\begin{eqnarray}
\begin{aligned}
e^\nu& \simeq1-\frac{r_g}{r}+\frac{r^2}{4t^2},
\label{nu1}\\
e^\lambda& \simeq1+\frac{r_g}{r}+\frac{r^2}{4t^2}.
\end{aligned}
\end{eqnarray}
However, if the condition $t\gg r_g$ is satisfied, then we
can write with the same accuracy in the entire range
$r_g<r<t$
\begin{eqnarray}
\begin{aligned}
e^\nu & \simeq1-\frac{r_g}{r}-\frac{1}{2}+\frac{t}{2\sqrt{t^2-r^2}},
\label{nu2}\\
e^\lambda & \simeq\left(1-\frac{r_g}{r}\right)^{-1}-\frac{1}{2}+\frac{t}{2\sqrt{t^2-r^2}}.
\end{aligned}
\end{eqnarray}
In the range (\ref{region1}) the sum of the solutions for the BH
and cosmology acts as a solution of the linearized
Einstein equations, while outside this range either the
BH metric or the cosmological one survives. The
latter assertion is based on the fact that if the metric
coefficients are formally represented as $1+\varepsilon$, then
outside the range (\ref{region1}) the coefficients of one of the
metrics are already $\varepsilon\sim1$, while the coefficients of the
other metric remain $\varepsilon\ll1$ and they may be neglected.

Let us call the ``radius of influence of BH'' the radius for which the two last terms in 
(\ref{nu1}) are of the same order,
\begin{equation}
r_{\rm infl}=(r_gt^2)^{1/3}.
\label{inflrad1}
\end{equation}
A general approach to estimating the radius of influence
based on a quasi-local mass was discussed in
Faraoni et al. (2015). The competition between local
attraction and cosmological expansion can be important
for the evolution and properties of astrophysical
objects; in particular, dark energy can affect the outer
regions of galaxy clusters (Bisnovatyi-Kogan 2015).
It is interesting to mention that the expression~(\ref{inflrad1}) can be also obtained 
from the condition that the BH mass becomes equal to the radiation mass in a
volume of radius $r$ (Eroshenko 2016).

When deriving Eqs.~(\ref{nu1}), we implicitly
assumed the test fluid approximation at $r\sim r_g$, while at $r\gg r_g$ the fluid density is close to the
cosmological density. This is indeed the result of
calculations of quasi-stationary accretion (Babichev
et al. 2004), where at $r\gtrsim10 r_g$ the fluid density already
differs very little from the density at infinity (see
Section 5 below).

It should be also mentioned that Eqs.~(\ref{nu2})
cannot be formally applied at very small radii close to $r_g$, 
because of a curvature singularity emerging at $r\simeq r_g$.
The reason for the appearance of the singularity is
the presence of cosmological corrections in Eqs.~(\ref{nu2}): as a result, the Schwarzschild-type coordinate singularity becomes
physical. This is an artefact originating from the use of the singular
coordinates for the BH metric. 
Therefore, for purposes that include curvature, the Schwarzschild metric (\ref{nuS})
should be used instead of (\ref{nu2}) at
$r\sim r_g$.

To find the form of the metric in the first approximation in the intermediate range~(\ref{region1}),
the linearized Einstein equations turned out
to be sufficient, as we have seen above.
However, in order to obtain the
density and velocity distributions one has to take into
account nonlinearities, i.e., to expand the Einstein
equations to the next orders in small parameters. 
Instead of working with the Einstein equations directly, in the next
section we will find for the fluid distribution and its velocity using
the equations of motion of the fluid.
The advantage of this  approach is that the Bianchi identifies automatically include the next to leading order in $r_g/r$ and $r/t$.


\section{Density and velocity distributions}

In this section we consider dynamics of a perfect fluid with the
energy-momentum tensor
\begin{equation}
\label{emt} T_{\mu\nu}=(\rho+p)u_\mu u_\nu - pg_{\mu\nu},\nonumber
\end{equation}
and the equation of state $p=\rho/3$ in the metric (\ref{nu2}). 
In this case, at large distances from the BH
this fluid sources the cosmological metric, while close to the BH the 
gravity of the black hole dominates and the backreaction of the fluid can be neglected.

The projection of the identities $T^{\mu\nu}_{\;\;\; ;\nu}=0$ onto
the 4-velocity $u_{\mu}T^{\mu\nu}_{\quad ;\nu}=0$ and their zero component
$T^{0\nu}_{\;\;\; ;\nu}=0$ read, respectively,
\begin{eqnarray}
&&\frac{\partial \rho}{\partial t}\left(\frac{1}{f_0}+\frac{u^2}{f_0f_1}\right)^{1/2}+u\frac{\partial \rho}{\partial r}
\label{big1}
\\
&+&\frac{4\rho}{3}\frac{f_1^{1/2}}{f_0^{1/2}}\left[\frac{\partial}{\partial t}\left(\frac{1}{f_1}+\frac{u^2}{f_1^2}\right)^{1/2}+\frac{1}{r^2}\frac{\partial }{\partial r}\left(\frac{f_0^{1/2}}{f_1^{1/2}}r^2u\right)\right]=0\nonumber
\end{eqnarray}
and
\begin{eqnarray}
&&\frac{f_1^{1/2}}{f_0^{1/2}}\frac{\partial}{\partial t}\left[\frac{f_0^{1/2}}{f_1^{1/2}}\rho\left(1+\frac{4u^2}{3f_1}\right)\right]
\label{big2}
\\
&+&\frac{4}{3r^2}\frac{f_1^{1/2}}{f_0^{1/2}}\frac{\partial}{\partial r}\left[r^2\frac{f_0^{1/2}}{f_1^{1/2}}\rho u\left(f_0+u^2\frac{f_0}
{f_1}\right)^{1/2}\right]
\nonumber
\\
&-&\frac{\rho}{2}\frac{\partial f_0}{\partial t}\left[\frac{1}{f_0}+\frac{4u^2}{3f_0f_1}\right]-\frac{\rho}{6f_1^2}\frac{\partial f_1}{\partial t}\left[f_1+4u^2\right]=0.\nonumber
\end{eqnarray}

Let us assume the range of radii (\ref{region1}).
We will look for the solution of Eqs.~(\ref{big1}) and (\ref{big2}) as an expansion in the small parameters
$r/t$ and $r_g/r$,
with certain asymptotics at the boundaries of the interval $r_g\ll r\ll t$.
We do not write down the resulting expressions here, since they are rather cumbersome.
First we note that
in the case of quasi-stationary accretion considered
by Babichev et al. (2004), the fluid density near a BH
is proportional to the density at infinity.
Therefore, we will assume (and below we check this  {\it a posteriori})
that in the problem under consideration the
time-varying density at a large distance from the BH
will enter into the solution as a common factor. 
Therefore, we make a conjecture that 
up to the lowest orders in small parameters, the density distribution can
then be written as
\begin{equation}
\rho=\frac{3}{32\pi t^2}\left(1+2\frac{r_g}{r}+\frac{r^2}{2t^2}+\alpha\frac{r_g}{t}\right)
\label{rhoappr}
\end{equation}
Note that we guessed the above expression in such a way that the asymptotic value of $\rho$ for large
$r$ corresponds to the cosmological solution (\ref{rhoinfty2}) expanded
in terms of the small parameters, 
while the asymptotic value at small $r$ were derived from the quasi-stationary
solution obtained by Babichev et al. (2004)
(see also Section 5). The last cross term in (\ref{rhoappr}) contains a free constant $\alpha$ and it can 
be written as a product of the two small parameters $\alpha (r_g/r)(r/t)$. Therefore, it does not
spoil the asymptotic values of $\rho$ at both large and small $r$.

Let us now guess the expression for the velocity $u$
satisfying Eqs.~(\ref{big1}) and (\ref{big2}) with density given by~(\ref{rhoappr}). 
Similarly to the expression for $\rho$, we will
write the velocity as an expansion up to the second
order in $r_g/r$ and $r/t$ by taking into account the
asymptotics (\ref{ucosm}) and the quasi-stationary solution
given in (Babichev et al. 2004):
\begin{equation}
u= \frac{r}{2t}-\frac{3^{3/2}}{2}\frac{r_g^2}{r^2}+\beta_1\frac{r^2}{t^2}+\beta_2\frac{r_g}{t}.
\label{uapprset}
\end{equation}
Substituting (\ref{rhoappr}), (\ref{uapprset}) and Eqs.~(\ref{nu1}) for the
metric into the expansion of Eqs.~(\ref{big1}) and (\ref{big2}) in
terms of the small parameters, we find that all terms
of the first order in $r/t$ and $r_g/r$ cancel out when
choosing $\beta_1=0$ and $\beta_2=3/4$. 
Moreover, the last two
terms in the expansion (\ref{rhoappr}) lead in Eqs.~(\ref{big1}) and (\ref{big2})
to terms $\sim r_g/t$ and $\sim(r/t)^2$, respectively, which are
higher order with respect to the leading terms $r/t$
and $r_g/r$; therefore, they may be discarded. 

It is important to stress here that the leading terms in Eqs.~(\ref{big1}) and (\ref{big2}) are determined not only by the 
order of small parameters, but also by the number of derivatives with respect to $r$ and $t$. 

Finally, In the range $r_g\ll r\ll t$ 
the approximate solutions for the density and
velocity read
\begin{eqnarray}
\rho& =& \frac{3}{32\pi t^2}\left(1+\frac{2 r_g}{r}\right),
\label{rhoappr2}\\
u &=& -\frac{3^{3/2}r_g^2}{2r^2}+\frac{3}{4}\frac{r_g}{t}+\frac{r}{2t}.
\label{uappr2}
\end{eqnarray}
These expressions, along with the metric (\ref{nu2}), are the main result of this paper.

For the density (\ref{rhoappr2}) it turns out to be sufficient in our approximation 
to choose simply the sum of the cosmological density and
density distribution in the quasi-stationary accretion.
In contrast, Eq.~(\ref{uappr2}) for the velocity contains---apart from the sum of the corresponding velocities---the extra
term $\propto r_g/t$ . 
Notice the positive coefficient in front of this term. 
This may seem surprising, since it means that for large enough radii $r\gg \sqrt{r_g t}$, 
(when the second term in (\ref{uappr2}) greater than the first one)
the BH repulse matter instead of attracting it. 
This behaviour, however, is explained by the fact that there
is a density excess around the BH, leading to non-zero pressure, which tends to drive away the fluid from the BH.

In the above calculations the BH
mass could be assumed to be constant in all equations,
except for Eqs.~(\ref{G01}) and (\ref{G01L}) describing the flow onto
the BH. In these equations, when differentiating with
respect to time, one should take into account the fact
that $\dot r_g\neq0$. We will consider the effect of the flow
onto the BH in the approximation of quasi-stationary
accretion in the next section.

From (\ref{uappr2}) we find the flow ``turning point'' defined as $u=0$:
\begin{equation}
r_{\rm turn}\simeq3^{1/2}(r_g^2t)^{1/3}.\nonumber
\label{turnrad1}
\end{equation}
This turning point recedes from the BH with a speed
$\dot r_{\rm turn}\simeq 3^{-1/2}r_g^{2/3}t^{-2/3}/3$. Note that the turning point
does not coincide with the radius of influence defined
above (\ref{inflrad1}). The characteristic scales are arranged as
follows:
\begin{equation}
r_g\ll r_{\rm turn}\ll\sqrt{r_gt}\ll r_{\rm infl}\ll t.\nonumber
\label{order}
\end{equation}


\section{QUASI-STATIONARY ACCRETION AT SMALL RADII}
\label{quassec}

At $r\ll r_{\rm turn}$ the solution of Babichev et al.
(2004) with the flow fixed at the critical point is applicable
with a good accuracy. Indeed, at $r\ll r_{\rm turn}$
in (\ref{uappr2}) the first term is dominant and the characteristic
accretion time of a fluid element from radius $r$ is
\begin{equation*}
\frac{r}{v}\sim\frac{r^3}{r_g^2}\ll t.
\end{equation*}
Therefore, the formalism from Babichev et al. (2004)
describes quasi-stationary accretion under which the
density at infinity is equal to the evolving cosmological
density. 
In this case, it is safe to neglect the time
derivatives and the terms of order $r^2/t^2$ in Eqs.~(\ref{big1})
and (\ref{big2}) and to set $f_0=f_1=f$. Whereupon the equations
take a simple form and give the following two
integrals of motion:
\begin{eqnarray}
u x^2\rho^{3/4} & =& C_1,
\label{c1eq}\\
u x^2\rho(f+u^2)^{1/2} &=&C_2,
\label{c2eq}
\end{eqnarray}
where $x\equiv r/r_g$, while $C_1$ and $C_2$ change slowly with time and depend on $\rho_{\infty}(t)$.

\begin{figure}[t]
\begin{center}
\includegraphics[angle=0,width=0.49\textwidth]{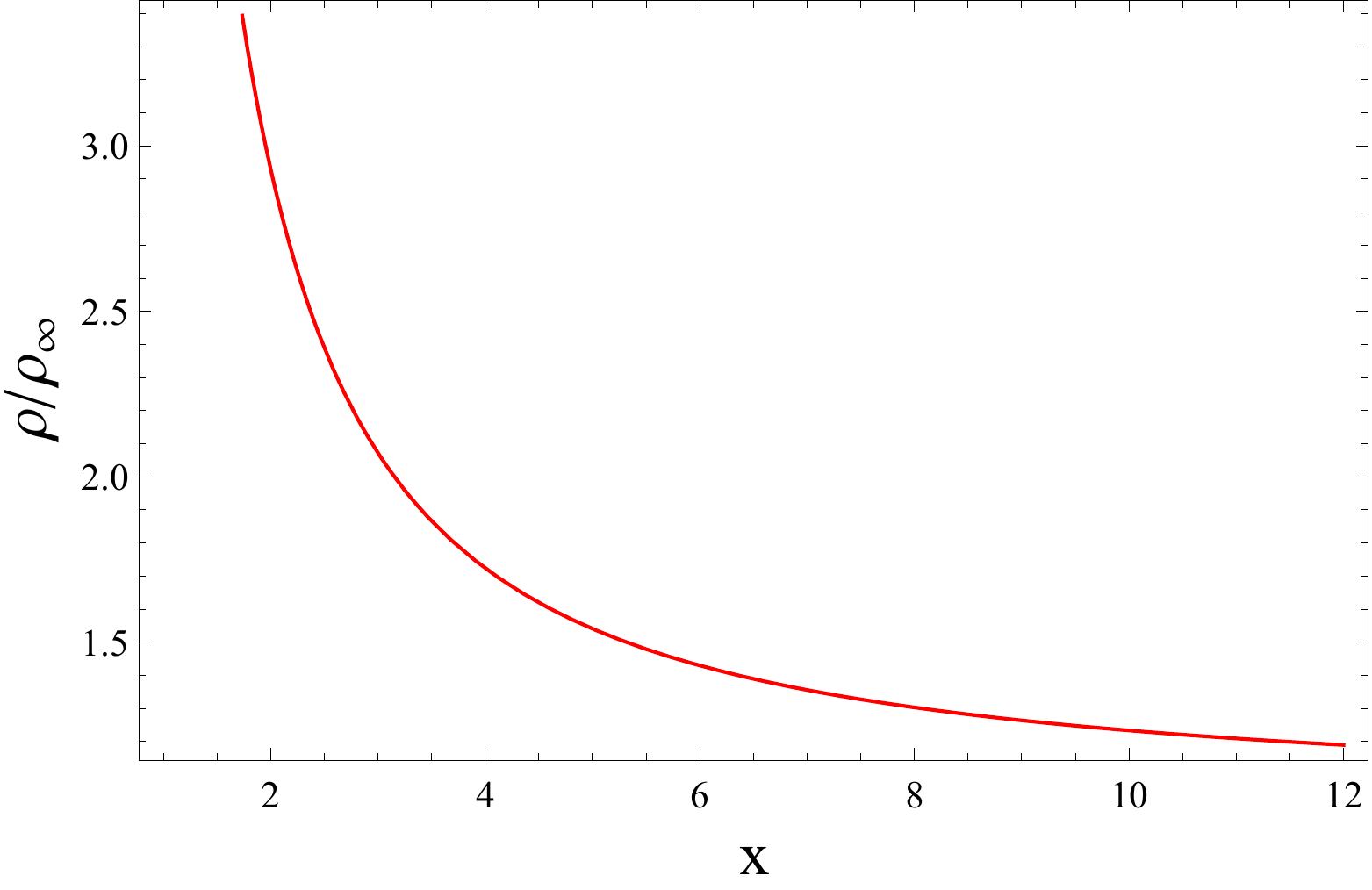}
\end{center}
\caption{Thermalized photon gas density near a black hole versus dimensionless radial variable $x=r/r_g$ in the
approximation of quasi-stationary accretion.} \label{grrho}
\end{figure}

The critical point of the fluid flow can be found by
a standard method (see, e.g., Michel 1972; Babichev
et al. 2005, 2013):
\begin{equation*}
x_*=\frac{3}{2}, \quad u_*=\frac{1}{\sqrt{6}}.
\end{equation*}
The density distribution is found from a cubic
equation that can be derived from (\ref{c1eq}) and (\ref{c2eq}).
As the result, the density distribution for quasi-stationary
accretion (at $r\ll r_{\rm turn}$) reads
\begin{equation}
 \label{sol1}
 \rho=\frac{81}{16}\frac{\rho_{\infty}(t)}{\omega^2 x^4},
\end{equation}
where
\begin{equation*}
 \omega=\left\{ \begin{array}{ll}
 \cos\left(\frac{\phi}{3}+\frac{2\pi}{3}\right),& 1\leq x\leq3/2,\\
 \cos\left(\frac{\phi}{3}-\frac{2\pi}{3}\right),& x>3/2,
\end{array} \right.
\end{equation*}
\begin{equation*}
 \phi=\arccos\left[\frac{27}{4 x^2}\left(1-\frac{1}{x}\right)\right].
\end{equation*}
This solution was obtained in a different form by
Babichev et al. (2004). The distribution (\ref{sol1}) is shown
in Fig.~1. 
At large distances $x\gg 6$ the solution (\ref{sol1})
has the asymptotic behaviour as follows,
\begin{equation}
 \label{sol2}
 \rho\simeq\rho_{\infty}(t)\left(1+\frac{2 r_g}{r}\right),\nonumber
\end{equation}
which coincides with (\ref{rhoappr2}) if we take into account (\ref{rhoinfty2})
at $r\ll t$.

The rate of accretion onto the BH is (Babichev
et al. 2004)
\begin{equation}
 \label{evol}
 \dot{M}=A \,\pi\, r_g^2 [\rho_{\infty}+p(\rho_{\infty})],\nonumber
\end{equation}
where $A=2\times3^{3/2}$ in the case under consideration (i.e. the equation of state is that for radiation).
If the density at infinity changes as the cosmological
density $\rho_{\infty}(t)=3/(32\pi t^2)$ and if the approximation
of quasi-stationary accretion is valid, then the BH
mass evolves as
\begin{equation*}
M=\frac{M_i}{1-\frac{3^{3/2}M_i}{t_i}\left(1-\frac{t_i}{t}\right)},
\end{equation*}
where $M_i$ is the mass of the primordial BHat the time
of its formation $t_i$. Thus, if the BH is formed with the
horizon radius $2M_i$, which is much smaller than the cosmological
horizon $2t_i$, then under further accretion the
growth of the mass is negligible, $M\approx const$. This is
a well-known result discussed in Carr et al. (2010).

The velocity of the flow $u$ is found from relation
(\ref{c1eq}) in the form
\begin{equation}
u=-\frac{3^{3/2}}{2}\left(\frac{\rho}{\rho_{\infty}}\right)^{3/4}\frac{1}{x^2}.
\label{fluxu}
\end{equation}
This expression for the velocity in the quasi-stationary
region was used to fix the asymptotics in (\ref{uapprset}).

\begin{figure}[t]
\begin{center}
\includegraphics[angle=0,width=0.49\textwidth]{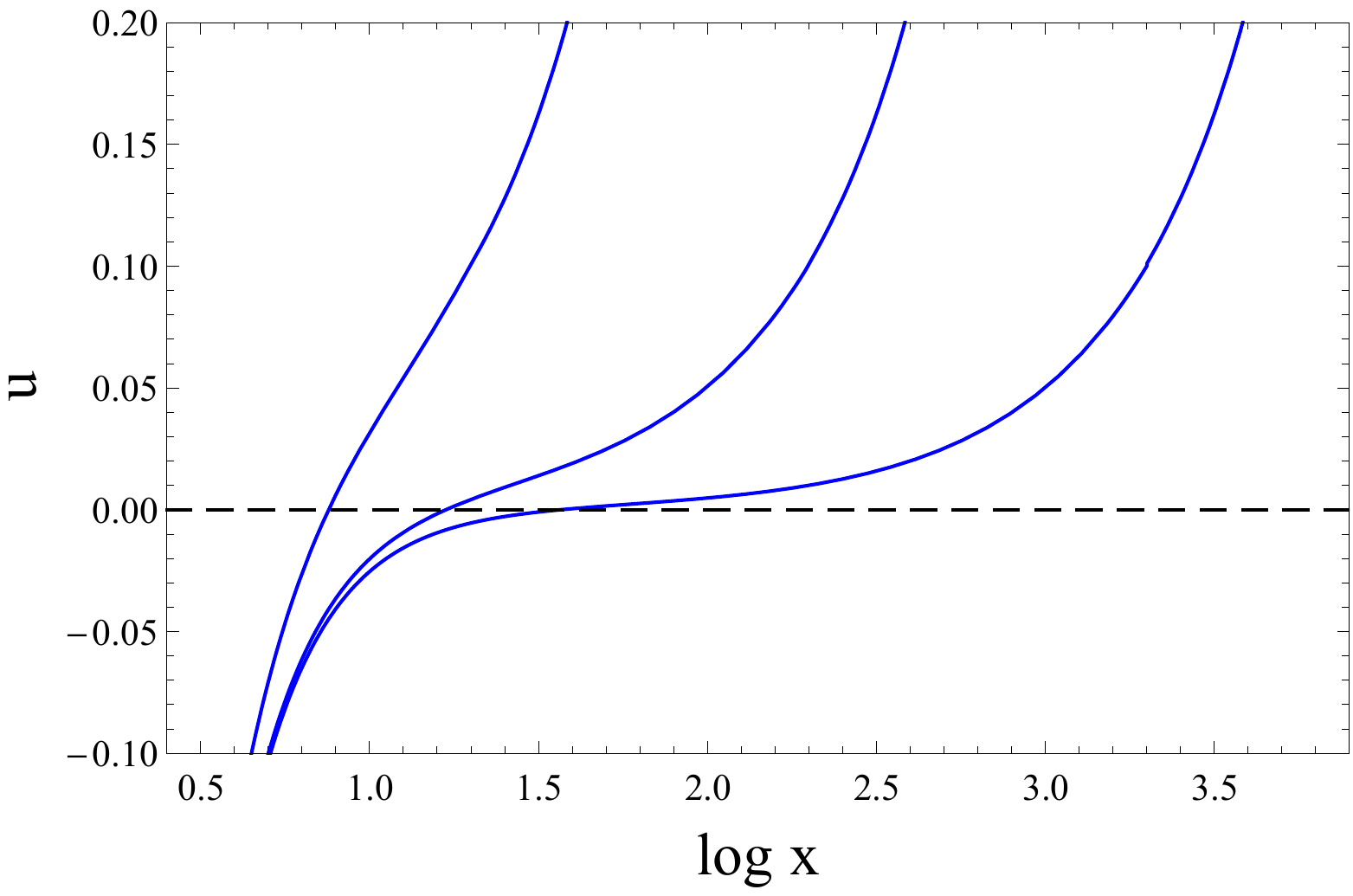}
\end{center}
\caption{Flow velocity $u$ versus dimensionless radial variable $x=r/r_g$ for $t/r_g=10^2$, $10^3$, and $10^4$ (from left to
right).} \label{gru}
\end{figure}

The flow velocity calculated from Eqs.~(\ref{uappr2}), (\ref{fluxu}),
and (\ref{ucosm}) (the last two formulas give the asymptotic
form of $u$ near the BH horizon and the cosmological
horizon) is shown in Fig.~2. We see that the turning
point $r_{\rm turn}$ moves to larger radii as time increases, as
we expected.

For practical calculations it is convenient to have
the formulas that express the matter density and velocity
in the entire range $r_g< r<t$. To do this we write
Eq.~(\ref{sol1}) as $\rho=k^2(x)\rho_{\infty}(t)$, where we defined the function $k(x)$ as
\begin{equation}
 \label{vark1}
k(x)=\frac{9}{4\omega x^2}.\nonumber
\end{equation}
Then, Eq.~(\ref{fluxu}) takes the form
\begin{equation}
u=-\frac{(3k)^{3/2}}{2x^2}.\nonumber
\label{fluxu2}
\end{equation}
The universal expressions, which have correct asymptotic behaviour
at the boundaries of the interval $r_g< r<t$ and which reproduce 
the approximate solution (\ref{rhoappr2}) and (\ref{uappr2}) in the
intermediate range $r_g\ll r\ll t$, can be written as follows,
\begin{eqnarray}
\begin{split}
\rho&=\frac{3 k^2(x)}{8\pi \left(t+\sqrt{t^2-r^2}\right)^2},
\label{rhoappr3} \\
u&=-\frac{(3k)^{3/2}}{2x^2}+\frac{3}{4}\frac{r_g}{t}+\frac{r}{t+\sqrt{t^2-r^2}}.
\end{split}
\end{eqnarray}

\begin{figure}[t]
\begin{center}
\includegraphics[angle=0,width=0.49\textwidth]{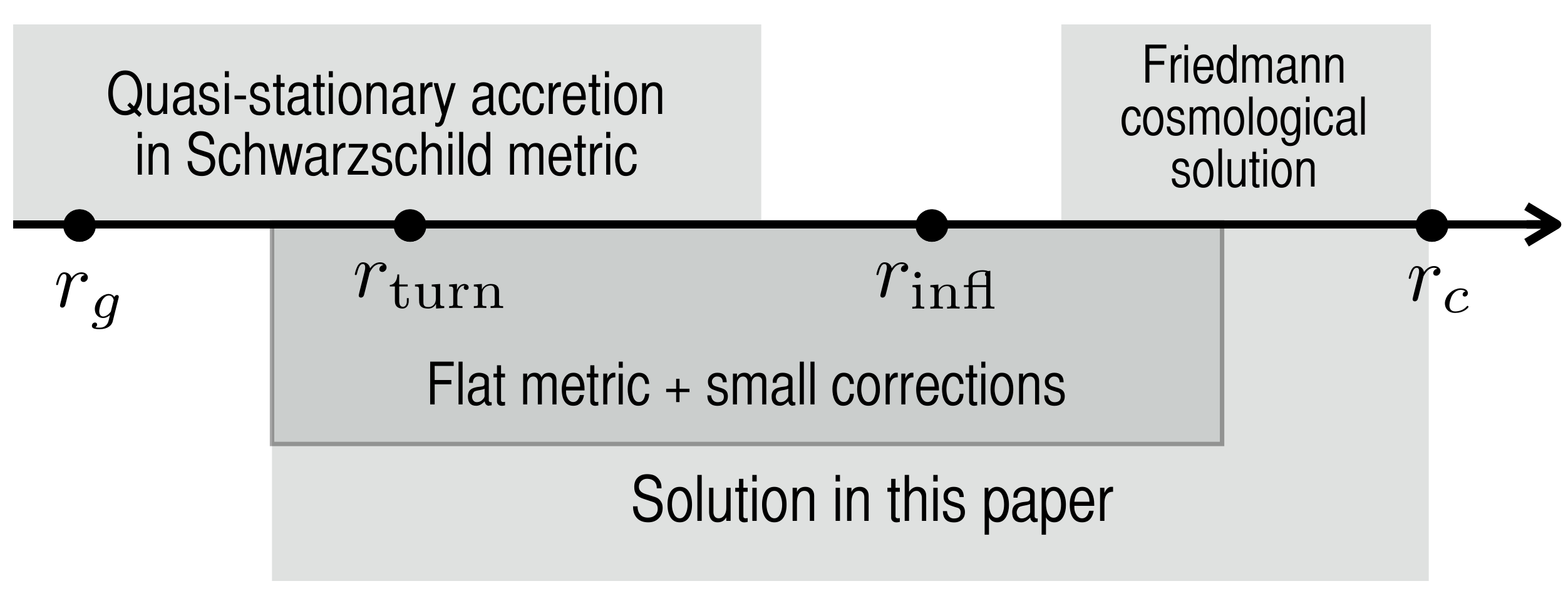}
\end{center}
\caption{Ranges of validity of quasi-stationary accretion, the Friedmann homogeneous cosmological solution, and the solution
found in this paper. The figure shows: the BH radius $r_g$; the turning point $r_{\rm turn}\sim (r_g^2 t)^{1/3}$, where the radial fluid velocity is
zero; the radius of influence of the BH $r_{\rm infl}\sim (r_g t^2)^{1/3}$, where the corrections to the flat metric from the BH are comparable
to the corrections from cosmology; the cosmological horizon $r_c\sim t$. The solution obtained in this paper (bottom gray area) is given by Eqs.~(\ref{EF2}), (\ref{nu2}) for the metric and by Eqs.~(\ref{rhoappr3}) for the density and velocity, respectively.} \label{pic}
\end{figure}

Figure~3 shows the ranges of validity of our solution
in comparison with the solution for quasistationary
accretion (Babichev et al. 2004)
and the Friedmann cosmological solution for a homogeneous Universe.

\vspace{1.5\baselineskip}

\section{CONCLUSION}

In this paper we presented an approximate solution
for the BH metric along with the density and velocity distributions of matter in an expanding radiation dominated universe.
The metric in our solution is given by Eqs.~(\ref{EF2}), (\ref{nu2}), while the density and velocity distributions
are given by Eqs.~(\ref{rhoappr3}). 
For distances $r_g\ll r \ll (r_g t^2)^{1/3}$ our solution matches the solution
for stationary accretion (Babichev et al. 2004),
while at large $r$ the obtained solution asymptotes the FLRW solution for a homogeneous universe.  
The obtained expressions for the velocity and density distributions, Eqs.~(\ref{rhoappr3}), are valid for
all $r$ between the BH horizon and the cosmological
horizon, $r_g<r<t$.
On the other hand, Eqs.~(\ref{nu2})
for the metric coefficients are valid for $r_g \ll r < t $, i.e. one cannot apply these expression for very small $r\sim r_g$,
due to a divergence in the curvature at $r\simeq r_g$. 
At small distances, $r\ll (r_g t^2)^{1/3}$, the metric must be replaced
by the Schwarzschild metric (\ref{nuS}) (see
Fig.~3, where the ranges of validity of various solutions
are shown).

In the case of the formation of a primordial BH,
our solution corresponds to a steady state, realised when sufficiently
long time has elapsed after the  BH formation (which happened at $t\sim r_g$), so that the acoustic waves scatter
to infinity and the cosmological horizon expanded to
$t\gg r_g$.

Because of non-zero pressure, the density of the thermalized photon gas is almost constant at distances much
smaller than the radius of influence of the BH (\ref{inflrad1}), $r\ll r_{\rm infl}$.
The fluid flow changes its direction (the point where $u=0$)
at $r=r_{\rm turn}\ll r_{\rm infl}$.
Our calculations show that for $t\gg r_g$ the photon
gas accretion onto the BH can be calculated by the
method developed by Babichev et al. (2004), with
the density at infinity replaced by the evolving
cosmological density. The accretion in the nearby
region occurs on short time scales, so that the density
in the nearby region has time to be adjusted to the
changing cosmological density.

It is interesting to compare our solution with the
well-known solution by McVittie (1933). The main
advantage of our solution is that we did not assume
zero flow onto the BH, in contrast to the McVittie solution.
We found the velocity and density distributions
by requiring that our solution asymptotes the quasi-stationary for small radii,
and that it assumes the cosmological FLRW solution at large distances.
In our solution the surrounding radiation flows onto the BH, creating a non-zero flux. 
In this respect our solution is physical and can be used for
practical applications, in particular, for physics of primordial BHs
and for the evolution of dark matter clumps at the
radiation-dominated stage. 
Unfortunately, similarly to
McVittie solution, the metric found in this paper
contains a divergence on the BH horizon. 
We hope to address this issue in future, using, e.g.,
a method similar to that considered in Babichev et al. (2012), where the inverse effect of matter was
taken into account for regions close to the BH horizon.


\section*{ACKNOWLEDGMENTS}

E. Babichev is grateful to V. Faraoni for the
useful discussions. The authors acknowledge support
from PRC CNRS/RFBR (20182020) no. 1985
``Gravite modifie e et trous noirs: signatures experimentales et mode les consistants'' and RFBR 18-52-15001 NCNI\_a (Russia); and from the research program ``Programme national de cosmologie et galaxies''
of the CNRS/INSU, France.


\end{document}